\documentclass[10pt]{article}
\usepackage{graphicx}
\usepackage{graphicx,graphics}
\begin{document}
\date{}

\title{Particle Creation by a Moving Boundary with Robin Boundary Condition}

\author{B. Mintz\footnote{mintz@if.ufrj.br} , C. Farina\footnote{farina@if.ufrj.br} ,
P.A. Maia Neto\footnote{pamn@if.ufrj.br} $\,$ and R.B. Rodrigues\footnote{robson@if.ufrj.br}\\
\\
Instituto de F\'\i sica - Universidade Federal do Rio de Janeiro\\
Caixa Postal 68528 - CEP 21941-972, Rio de Janeiro, Brasil.\\
\\}
\bigskip
\maketitle
\begin{abstract}
{We consider a massless scalar field in 1+1 dimensions satisfying a Robin boundary condition (BC) at a non-relativistic moving boundary. 
We derive a Bogoliubov transformation between input and output bosonic field operators, which allows us 
to calculate the  spectral distribution of created particles. The cases of Dirichlet and Neumann BC may be obtained 
from 
our result as limiting cases. These two limits yield the same spectrum, which turns out to be an upper bound 
for the spectra derived for Robin BC. We show that the particle emission effect can be considerably reduced 
(with respect to the Dirichlet/Neumann case) by selecting a particular value for the oscillation frequency of 
the boundary position.  
}
\end{abstract}



\section{Introduction}

Moving bodies experience fundamental energy damping \cite{damping} \cite{FordVilenkin}
and decoherence \cite{decoherence}  mechanisms
 due to the scattering of vacuum field fluctuations.
The damping is accompanied by the emission of particles (photons in the case of the electromagnetic field) 
\cite{photons}, thus
conserving the total energy of the body-plus-field system \cite{Lambrecht}\cite{energy}.  
This dynamical (or nonstationary) Casimir effect
has been analyzed for a variety of three-dimensional geometries, including 
parallel plane plates \cite{Mundarain}, cylindrical waveguides \cite{waveguide},
and  rectangular \cite{closed}, cylindrical \cite{cylindrical-cav} and spherical cavities \cite{spherical}.  
It also depends on the details of the coupling between the field and the
body, which can usually be cast in the form of boundary conditions (BC) for the field. 
Of particular theoretical relevance is the Robin BC, which continuously interpolates the Dirichlet and Neumann BC. 
For a massless scalar field in 1+1 dimensions, it reads
\begin{equation}\label{Robin}
 \frac{\partial\phi}{\partial x}(t,x_0)=\frac{1}{\beta}\,\phi(t,x_0),
\end{equation} 
where $x_0$ is the position of the boundary. 
The positive parameter $\beta$ represents a time scale (we take $c=1$)
associated to the time delay 
(or phase shift) characteristic of reflection at the Robin boundary \cite{Robin-Forca}, which
 can be interpreted in terms of a simple mechanical model \cite{Chen}.
According to eq.~(\ref{Robin}), Dirichlet and Neumann BC are obtained as the
 limiting cases $\beta\rightarrow 0$ and
$\beta\rightarrow  \infty,$ respectively.  

We consider
a semi-infinite  slab (extending from 
$-\infty$ to $x=\delta q(t)$)
  following a prescribed nonrelativistic motion, with the Robin boundary at $\delta q(t).$ 
We have recently
computed the dynamical Casimir force on the slab, which contains  dissipative as well as dispersive components \cite{Robin-Forca}. 
In this paper, we will analyze in detail the particle creation effect  and compute 
the corresponding spectral distribution. 

\section{Input-Output Bogoliubov Transformation}

In the instantaneously
co-moving frame
the  massless scalar field  satisfies
\begin{equation}
 \frac{\partial\phi'}{\partial x'}|_{\rm bound}=\frac{1}{\beta}\,\phi'|_{\rm bound},
\end{equation} 
Neglecting terms of the order of $[\delta {\dot q}(t)]^2,$ we find, in the laboratory frame,
\begin{equation}\label{Robin1}
\left[ \frac{\partial}{\partial x}+
\delta {\dot q}(t) 
 \frac{\partial}{\partial t}\right] \phi(t,\delta {q}(t))=
\frac{1}{\beta}\,\phi(t,\delta {q}(t)).
\end{equation} 

We assume that final position coincides with initial one, which is taken at $x=0.$ Hence 
\begin{equation}
 \lim_{t\rightarrow\pm\infty}\delta q(t)=0.
\end{equation}
Jointly with the nonrelativistic approximation, this condition implies that 
$\delta q(t)$ is much smaller than the wavelengths $\lambda$ of the created particles.  
In fact, we will show that the frequencies of the particles are bounded by the mechanical 
frequencies $\omega_0:$ $\omega =2\pi/\lambda\le \omega_0.$ Since $\omega_0 \delta q \sim \delta{\dot q}\ll 1,$
we have $\delta q(t) \ll \lambda.$

Thus, we may analyze eq.~ (\ref{Robin1}) by expanding up
to first order in $\delta q$ and its derivatives. This amounts to calculate the 
effect of the motion as a small perturbation \cite{FordVilenkin}:
\begin{equation}
 \phi(t,x)=\phi_0(t,x)+\delta\phi(t,x),
\end{equation}
where the unperturbed field
$\phi_0$ corresponds to 
a solution with a static boundary 
at $x=0.$
The first-order field $\delta \phi$ then satisfies the following BC at $x=0:$
\begin{equation}\label{first-order}
\frac{\partial\delta\phi}{\partial x}(t,0)-
\frac{1}{\beta}\delta\phi(t,0)=
\delta q(t)\left[\frac{1}{\beta}
\frac{\partial\phi_0}{\partial x}(t,0)-
\frac{\partial^2\phi_0}{\partial x^2}(t,0)\right] - 
\delta \dot q(t)
\frac{\partial\phi_0}{\partial t}(t,0)\, .
\end{equation}

It is convenient to use the Fourier representation
\[
\Phi(\omega,x)=\int dt e^{i\omega t} \phi(t,x).
\]
The unperturbed field satisfies the Robin BC at $x=0.$ Its 
normal mode expansion  for $x>0$ is given by 
\begin{equation}\label{Phi0}
\Phi_0(\omega,x)=N(\omega)
[\sin(\omega x) + \omega\beta\cos(\omega x)]
\left[\Theta(\omega)a(\omega) - \Theta(-\omega)a(-\omega)^{\dagger}\right],
\end{equation}
with 
\[
N(\omega)=\sqrt{\frac{4\pi}{|\omega|(1 + \beta^2\omega^2)}}
\]
and $\Theta(x)$ denoting Heaviside step function.
The bosonic operators $a(\omega)$ and $a(\omega)^{\dagger}$
satisfy the commutation relation
\begin{equation}
[a(\omega),\,a(\omega')^\dagger]=2\pi\,\delta(\omega-\omega').
\end{equation}

To solve eq.~(\ref{first-order}) for 
$\delta\Phi(\omega,x)$ in terms of $\Phi_0(\omega,0)$
(with
$x>0$), we  use suitably defined Green functions, obeying 
the differential equation
\begin{equation}\label{Green}
 \left(\frac{\partial^2}{\partial x^2} + \omega^2 \right)G(\omega,x,x')=\delta(x-x').
\end{equation}
>From Green's theorem, we find 
\begin{equation}\label{Green2}
 \delta\Phi(\omega,x')=-\delta\Phi(\omega,0)\frac{\partial }{\partial x}G(\omega,0,x') + 
G(\omega,0,x')\frac{\partial }{\partial x}\delta\Phi(\omega,0).
\end{equation}

This result is more easily combined with eq.~(\ref{first-order}) if we select
a solution $G_{\rm R}(\omega,x,x')$ 
of eq.~(\ref{Green}) satisfying the Robin BC at $x=0.$
With this Robin Green function, we immediately obtain the first-order field from 
eq.~(\ref{Green2}) in terms of the 
BC satisfied by $\delta \Phi(\omega,x)$ as given by the Fourier transform of eq.~(\ref{first-order}). 
Then, the complete field is written as 
\begin{equation}\label{phi}
 \Phi(\omega,x)=\Phi_{0}(\omega,x) + G_{\rm R}(\omega,0,x)\left[\frac{\partial }{\partial x}\delta\Phi(\omega,0) - \frac{
\delta\Phi(\omega,0)}{\beta}\right],
\end{equation}
with 
\begin{equation}\label{BC-on-PHI}
\frac{\partial }{\partial x}\delta\Phi(\omega,0) - \frac{
\delta\Phi(\omega,0)}{\beta}=\frac{1}{\beta}\,\int\frac{d\omega'}{2\pi}\left[\frac{\partial\Phi_{0}}{\partial x}(\omega,0) + \omega\omega'\Phi_{0}(\omega,0) \right]\delta Q(\omega - \omega'),
\end{equation}
where $\delta Q(\omega)$ is the Fourier transform of $\delta q(t).$

If we replace $G_{\rm R}$ in eq.~(\ref{phi}) by the retarded Robin Green function, given by
\begin{equation}\label{GreenRet}
 G_{\rm R}^{\rm ret}(\omega,0,x)=\frac{\beta}{1-i\beta\omega}e^{i\omega x},
\end{equation}
then the zeroth-order field $\Phi_0(\omega,x)$ corresponds to the input field $\Phi_{\rm in}(\omega,x),$ with
\[
\phi_{\rm in}(t,x)= \lim_{t\rightarrow -\infty} \phi(t,x).
\]
On the other hand, when taking the advanced Robin Green function, given by 
\begin{equation}\label{GreenAdv}
 G_{\rm R}^{\rm adv}(\omega,0,x)=\frac{\beta}{1+i\beta\omega}e^{-i\omega x},
\end{equation}
$\Phi_0(\omega,x)$ corresponds to the output field $\Phi_{\rm out}(\omega,x)$ 
($\phi_{\rm out}(t,x)= \lim_{t\rightarrow \infty} \phi(t,x)$). 
By combining these two possibilities, we find the relation between output and input fields:
\begin{equation}\label{Phi:OUT-IN}
 \Phi_{\rm out}(\omega,x)=\Phi_{\rm in}(\omega,x) + 
\left[
G_{\rm R}^{\rm ret}(\omega,0,x)-
G_{\rm R}^{\rm adv}(\omega,0,x)
\right]
\bigg[\frac{\partial }{\partial x}\delta\Phi(\omega,0) - 
\frac{\delta\Phi(\omega,0)}{\beta}\bigg]
\end{equation}
The final result is obtained by inserting eqs.~(\ref{BC-on-PHI})
(with $\Phi_0$ replaced by $\Phi_{\rm in}$ sin\-ce we neglect terms of second order),
(\ref{GreenRet}) and (\ref{GreenAdv}) into the rhs of eq.~(\ref{Phi:OUT-IN}). 
Further physical insight
is gained if we write 
the input-output relation in terms of
the annihilation operators $a_{\rm in},$ $a_{\rm out}$ and their Hermitian conjugates, by
combining eqs.~(\ref{Phi0}) and (\ref{Phi:OUT-IN}).  
The resulting input-output relation has the form of a Bogoliubov transformation:
\begin{eqnarray}\label{Bogoliubov}
 a_{\rm out}(\omega) & = &  a_{\rm in}(\omega) + \frac{2i\sqrt{\omega}}{\sqrt{1+\beta^2\omega^2}}
\int\frac{d\omega'}{2\pi}\frac{1+\beta^2\omega\omega'}{\sqrt{1+\beta^2\omega'^2}}
\sqrt{|\omega'|} \\
 & &\times\bigl[\theta(\omega')a_{\rm in}(\omega')-\theta(-\omega')a_{\rm in}(-\omega')^\dagger \bigr]\delta Q(\omega-\omega').\nonumber
\end{eqnarray}

Since the output annihilation operator is contaminated by the input creation operator, the 
input vacuum state $|0_{\rm in} \rangle$ is not a vacuum state with respect to the output operators. In the next section,
we compute the  resulting particle creation effect. 

\section{Frequency spectrum}

The number of particles created with frequencies between $\omega$ and $\omega + d\omega$ ($\omega\ge 0$) 
is 
\begin{equation}\label{spectrum1}
 \frac{dN}{d\omega}(\omega)\,d\omega = \langle0_{\rm in}| 
a_{\rm out}(\omega)^\dagger a_{\rm out}(\omega)| 0_{\rm in}\rangle \,\frac{d\omega}{2\pi} 
\end{equation}
The spectrum is obtained by inserting eq.~(\ref{Bogoliubov}) into (\ref{spectrum1}):
\begin{equation}\label{Spec}
 \frac{dN}{d\omega}(\omega) = \frac{2\omega}{\pi(1+\beta^2\omega^2)}\int_0^\infty\,\frac{d\omega'}{2\pi}\frac{\omega'[1-\beta^2\omega\omega']^2}{1+\beta^2\omega'^2}[\delta Q(\omega+\omega')]^2
\end{equation}

To single out the effect of a given Fourier component of the motion, we take 
\[
\delta q(t)=\delta q_0\cos(\omega_0 t)e^{-|t|/T}
\] 
with $\omega_0 T \gg 1.$ In this case, $\delta Q(\omega)$ corresponds to two very narrow peaks around
$\omega=\pm \omega_0,$ so that we may take the approximation
\begin{equation}\label{Qomega}
|\delta Q(\omega)|^2
 \approx \frac{\pi}{2} \,\delta q_0^2\, T \left[ \delta(\omega-\omega_0) + \delta(\omega+\omega_0)\right].
\end{equation}
Inserting this equation into (\ref{Spec}), we find
\begin{equation}\label{resposta}
 \frac{dN}{d\omega}(\omega)=\frac{\delta q_0^2\, T}{2\pi}\,\omega(\omega_0-\omega)\,
\frac{[1-\beta^2\omega(\omega_0-\omega)]^2}{(1+\beta^2\omega^2)[1+\beta^2(\omega_0-\omega)^2]}\, \Theta(\omega_0-\omega).
\end{equation}
Note that the spectrum vanishes for $\omega>\omega_0:$ no particle is created with frequency larger than the 
mechanical frequency. Field modes at higher frequencies are not excited by the  motion, which is slow
in the time scale corresponding to such frequencies (quasi-static
regime). This important property confirms the consistency of our perturbation approach, with its expansion 
in $\delta q/\lambda.$

\begin{figure}[!hbt]
\begin{center}
\includegraphics[width=7cm]{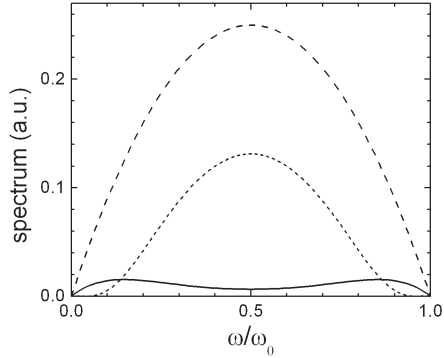}
\caption{\footnotesize Spectral distribution of the emitted particles 
$dN/d\omega.$ For the horizontal scale, we divide the frequencies by the 
  mechanical frequency $\omega_0.$ Dashed line: $\beta =0$ (Dirichlet case), solid line: $\beta\omega_0=1.7,$ 
dotted line  $\beta\omega_0=5$. 
}
\end{center}
\label{Distr_Espectral}
\end{figure}

A second important general property of the spectrum given by eq.~(\ref{resposta}) is the symmetry around 
$\omega=\omega_0/2:$ the spectrum is invariant under the replacement $\omega\rightarrow \omega_0-\omega.$
This is a signature that the particles are created in pairs, with frequencies such that their sum equals 
$\omega_0.$ Hence, for each particle created at frequency $\omega,$ there is a `twin' particle 
created at frequency $\omega_0-\omega.$

Since  Robin BC interpolate  Dirichlet and Neumann ones, we may derive the spectra for 
these two cases by taking appropriate limits of
eq.~(\ref{resposta}). 
For Dirichlet BC, we find
\begin{equation}
 \left.\frac{dN}{d\omega}(\omega)\right|_{(D)}= 
\lim_{\beta\rightarrow0}\frac{dN}{d\omega}(\omega)= 
\frac{(\delta q_0)^2\,T}{2\pi}\,\omega(\omega_0-\omega)\, \Theta(\omega_0-\omega),
\end{equation}
in agreement with Ref.~\cite{Lambrecht}. For the Neumann BC ($\beta\rightarrow\infty$), we find the {\it same spectrum}, 
confirming the equivalence between Dirichlet and Neumann in the context of the dynamical Casimir effect
in 1+1 dimensions \cite{AlvesFarinaPAMN}.  

For intermediate values of $\beta,$ the spectrum is always {\it smaller} than in the Dirichlet case
for all values of $\omega.$ 
In fact, we may write the result of eq.~(\ref{resposta}) as 
\begin{equation}
\frac{dN}{d\omega}(\omega)= \eta\,\frac{dN_{(D)}}{d\omega}(\omega),
\end{equation}
where the {\it reduction factor} $\eta\le 1$ is a function of $\beta\omega_0$ and $\omega/\omega_0.$ 
The reduction may be more severe near $\omega=\omega_0/2,$ which is the spectrum maximum in the Dirichlet case,
for some values of $\beta \omega_0$.
Hence, the Robin spectrum may develop global maxima near $\omega=0$ and $\omega=\omega_0,$ as in the example 
shown in Fig.~1 (solid line), with $\beta\omega_0=1.7$. In this figure, we also plot the Dirichlet/Neumann spectrum
(dashed line) and the Robin spectrum for  $\beta\omega_0=5$ (dotted line).
As discussed above, all curves are symmetric with respect to 
$\omega=\omega_0/2.$

The areas below the curves shown in Fig.~1 correspond to the total number of created particles.
The figure already indicates that this number, to be discussed in the next section, may be considerably reduced (with respect to the Dirichlet case) for intermediate values of $\beta.$

\section{Particle Creation Rate}

The total number of created particles is given by 
\begin{equation}\label{N}
N=\int_0^{\omega_0}\frac{dN}{d\omega}(\omega)\,d\omega = \frac{\delta q_0^2\,T}{2\pi}\,\omega_0^3\,F(\beta \omega_0),
\end{equation}
with
\begin{equation}
F(\xi)= \frac{\xi[4\xi+\xi^3+12\arctan(\xi)]-6(2+\xi^2)\ln(1+\xi^2)}{6\xi^2(4+\xi^2)}
\end{equation}
As expected for an open geometry (with a continuum of field modes), $N$ is proportional to the time $T,$ so that 
the particle creation {\it rate} $R\equiv N/T$ is the physically meaningful quantity. For the Dirichlet case, we take 
$F(\xi\rightarrow 0)=1/6,$ and then
\begin{equation}\label{RD}
R_{(D)} = \frac{\delta q_0^2 \omega_0^3}{12\pi}.
\end{equation}
For the Neumann case, we find the same result for the creation rate, since the spectrum is the same.
Note that the rate {\it increases} with $\omega_0$ according to eq.~(\ref{RD}) and vanishes (as required) in 
the static limit $\omega_0=0.$ This could have been anticipated since the particle creation  is an effect 
of changing the BC nonadiabatically. However, for Robin BC 
 the rate is not a monotonic function of $\omega_0.$
In Fig.~2,  we plot the rate $R$ as a function of $\beta\omega_0$ (for a fixed $\beta$). 
$R$ decreases as $\omega_0$ varies from 
$1.3/\beta$ to the local minimum at $2.1/\beta.$   

\begin{figure}[!hbt]
\begin{center}
\includegraphics[width=8cm]{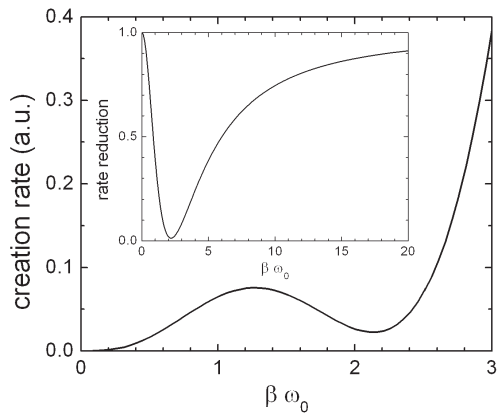}
\caption{\footnotesize Total particle creation rate as a function of mechanical frequency (in units of $1/\beta$).
Inset: ratio between creation rates for Robin and Dirichlet BC. }
\end{center}
\end{figure}

In the inset of Fig.~2, we plot the ratio $R/R_{(D)}=6 F(\beta\omega_0)\le 1$ as  a function of $\beta\omega_0.$ 
This ratio  represents the reduction of the Dirichlet creation rate for a finite $\beta.$ It only depends on the 
the dimensionless variable $\beta\omega_0,$ and goes asymptotically to one for $\beta\omega_0\gg 1,$
since the Neumann BC yields the same rate as the Dirichlet case. The reduction is maximum at 
$\beta\omega_0= 2.2$. At this point, 
the creation rate is reduced to $1.3\%$ of the Dirichlet value \footnote{When plotting the creation rate itself,
the effect seems  to be less impressive because the Dirichlet rate increases with $\omega_0.$}.

We may also calculate the radiated energy from these results. Thanks to the symmetry of the spectrum around 
$\omega=\omega_0/2,$ we have
\begin{equation}
E=\int_0^{\omega_0}\frac{dN}{d\omega}(\omega)\,\hbar\omega\,d\omega =\frac{ \hbar \omega_0}{2}\,N.
\end{equation}
Combining with eq.~(\ref{N}), we find
\begin{equation}\label{Ef}
E=\delta q_0^2\,T\,\hbar\omega_0^4\,F(\beta \omega_0)/(4\pi).
\end{equation}

This expression can be directly compared with the result for the Casimir  force we have recently 
reported \cite{Robin-Forca}. 
The force is written (in the Fourier domain) as ${\cal F}(\omega) = \chi(\omega) \delta Q(\omega)$
and its work on the slab is given in terms of the imaginary part of the susceptibility function $\chi(\omega)$:
\begin{equation}
 W=-\frac{1}{\pi}\int_0^\infty d\omega\,\omega\,
{\rm Im}\,\chi(\omega)\,\vert\delta Q(\omega)\vert^2.
\end{equation}
For the quasi-sinusoidal motion considered here, we find, using eq.~(\ref{Qomega}),
\begin{equation}\label{W}
 W = -\delta q_0^2\,T\,\omega_0\,{\rm Im}\,\chi(\omega_0)/2.
\end{equation}
The result for ${\rm Im}\,\chi(\omega_0)$ derived in Ref.~\cite{Robin-Forca} can be cast in
the form
\footnote{In Ref.~\cite{Robin-Forca}, a narrow plate is considered, rather than a slab. The  two sides of the plate provide identical (and independent) contributions to the force. Hence, to compare with the present situation, we 
divide the result of  \cite{Robin-Forca} by two.}
 ${\rm Im}\,\chi(\omega_0)=\hbar\omega_0^3F(\beta\omega_0)/(2\pi).$
Then, the comparison of eqs.~(\ref{Ef}) and (\ref{W}) yield $E=-W,$ so that 
the total radiated energy coincides with the negative of the work done on the slab by the 
Casimir force, as expected from energy conservation.

\section{Conclusion}

Dirichlet ($\beta\rightarrow 0$) and Neumann BC
($\beta\rightarrow \infty$)
 yield the same result for the 
spectrum of created particles. With the Robin BC, we are able to interpolate
continuously between these two cases. For intermediate values of $\beta,$ 
the spectrum is always smaller than the Dirichlet/Neumann case, for all 
values of frequency.  

In the range $1.2<\beta\omega_0<2.4$ the spectrum develops lateral peaks higher than the value at 
$\omega=\omega_0/2.$ This is also approximately the range in which the total creation rate 
(surprisingly)
decreases 
with $\omega_0.$ 
This  rate is reduced by up to $1.3\%$ of the Dirichlet/Neumann value, if the 
mechanical frequency is selected at $\omega_0=2.2/\beta.$ 
In other words, the coupling with the vacuum field state is considerably reduced if 
the slab oscillates at a frequency close to this value.

When considering the electromagnetic field  and a plane 
mirror moving along its normal direction, 
the BC in the ideal case of {\it perfect reflectors} may be decomposed into Dirichlet and 
Neumann BC for each orthogonal polarization \cite{JPA94}. 
In 3+1 dimensions, 
the effect with Neumann BC is considerably larger than with Dirichlet BC, and it would 
be interesting to investigate the continuous transition between these two limiting cases. 

Reflection by real metallic plates involve non-trivial frequency-dependent phase factors as in the case of 
Robin BC. The results of the present 
paper indicate that 
finite conductivity might yield a significant reduction of the  magnitude
of the dynamical Casimir effect.

This work was supported by the Brazilian agencies
 CNPq and FAPERJ. PAMN acknowledges Instituto do Mil\^enio de
Informa\c c\~ao Qu\^antica  for partial financial support. 


\end{document}